\documentclass{aa}
\usepackage{amsfonts,amsmath,units,wasysym,epsfig,graphicx,verbatim,color,subfigure,graphicx,bm,mathrsfs,lipsum,hyperref}
\usepackage[utf8]{inputenc}

\usepackage[normalem]{ulem}  

\begin{document}

\title{Speed of sound in dense matter and two families of compact stars}

\author{Silvia Traversi \inst{1,2} \and Prasanta Char \inst{2,3} \and Giuseppe Pagliara \inst{1,2} \and Alessandro Drago \inst{1,2}}

\institute{Dipartimento di Fisica e Scienze della Terra,  Università di Ferrara, Via Saragat 1, 44122 Ferrara, Italy 
\and INFN Sezione di Ferrara, Via Saragat 1, 44122 Ferrara, Italy 
\and Space  sciences,  Technologies  and  Astrophysics  Research  (STAR)  Institute,  Université de Liège, Bât. B5a, 4000 Liège, Belgium}

\abstract{
The existence of massive compact stars $(M\gtrsim 2.1 M_{\odot})$ implies that the
conformal limit of the speed of sound $c_s^2=1/3$ is violated if those stars have a crust of ordinary nuclear matter. Here we show that, if the most massive objects are strange quark stars, i.e. stars entirely composed of quarks, the conformal limit can be respected while observational limits on those objects are also satisfied.
By using astrophysical data associated with those massive stars, derived from electromagnetic and gravitational wave signals, we show, within a Bayesian analysis framework and by adopting a constant speed of sound equation of state, that the posterior distribution of $c_s^2$ is peaked around 0.3, and the maximum mass of the most probable equation of state is $\sim 2.13 M_{\odot}$. We discuss which new data would require a violation of the conformal limit even when considering strange quark stars, in particular we analyze the possibility that the maximum mass of compact stars is larger than $2.5M_{\odot}$, as it would be if the secondary component of GW190814 is a compact star and not a black hole. Finally, we discuss how the new data for PSR J0740+6620 obtained by the NICER collaboration compare with our analysis (not based on them) and with other possible interpretations.}

\maketitle

\section{Introduction}\label{intro}
After the discovery of neutron stars with masses $\sim 2M_{\odot}$ \citep{Demorest:2010bx, Antoniadis:2013pzd, Cromartie:2019kug} it became clear that the equation of state (EoS) of dense baryonic matter must be rather stiff to support such a large mass against gravitational collapse. 
The stiffness of the equation of state for nucleonic matter is regulated by the adiabatic index or equivalently by the speed of sound $c_s$ i.e. the derivative of pressure with respect to the energy density at fixed entropy. A remarkable result obtained in \citet{Bedaque:2014sqa} (see also the more recent \citep{Reed:2019ezm}) is that while it is expected 
that at asymptotically high density the speed of sound must respect the conformal limit of $c_s^2=1/3$ due to the QCD asymptotic freedom, at the densities reached in the core of neutron stars this bound seems to be violated: $c_s$ must increase to values significantly larger than the conformal bound and then it should decrease to reach asymptotically the conformal limit. In particular, if one assumes that compact stars have an external layer of ordinary nuclear matter up to a density $\rho_c\sim 2 \rho_0$ (where $\rho_0$ is the nuclear matter saturation density) and that at densities $\rho>\rho_c$ matter has a constant speed of sound $c_s^2 = 1/3$ one realizes that masses $\sim 2 M_\odot$ cannot be obtained \citep{Bedaque:2014sqa}. If one instead matches external layer and core at $\sim 1.1 \rho_0$, masses up to 2 $M_\odot$ can be obtained, but not larger than that \citep{Annala:2019puf}. 
While this need of violating the conformal limit stems from the request that the mass of a star with a ordinary crust exceeds $\sim 2 M_\odot$, larger violations are requested if conditions on the radii are also imposed. For instance, in order to obtain radii smaller than about 12 km for masses about $(1.4-1.5) M_\odot$ one can assume that a phase transition takes place at some density larger than $\rho_0$, obtaining stars having a core of quark matter separated, via a large energy density jump, from a nucleonic layer. Small radii can then be obtained in particular in the so-called twin stars model \citep{Alvarez-Castillo:2017qki,Christian:2019qer,Blaschke:2020qqj}, but only if $c_s$ is set to be very close to the causal limit $c_s=1$ \citep{Chamel:2012ea,Alford:2013aca,Alford:2015dpa}.
Notice that this inferred behaviour for $c_s$ is different from what happens at finite temperature and zero density matter, where lattice QCD has definitely established that $c_s$ is always below the conformal limit \citep{Karsch:2006sf}, a behaviour predicted also in several weak couplings and strong couplings theories, see \cite{Bedaque:2014sqa}.

There are two possible ways to solve the tension between the existence of massive stars
and the theoretical expectation 
to recover the high density conformal limit: 
\begin{itemize}

\item[(i)]
 there is some physical mechanism which is responsible for a rapid increase of $c_s$ at densities close to twice saturation density and which then "switches off" at larger densities. At those densities $c_s$ should decrease again below the conformal limit and then reach it asymptotically from below in agreement with pQCD calculations. There are several examples in the literature for this kind of explanation, see \cite{Hoyos:2016cob,Tews:2018kmu,Ma:2018qkg,Khaidukov:2018vkv,McLerran:2018hbz,Marczenko:2020wlc}.

\item[(ii)]
 the conformal limit is never violated because two types of stars exist, Hadronic Stars (HSs) and Strange Quark Stars (QSs) as described by the so-called two-families scenario \citep{Drago:2013fsa,Drago:2015cea,Drago:2017bnf,Char:2019wvo,DePietri:2019khb}. In that scenario the most massive stars are QSs and, as shown in the present paper, the conformal limit can be respected even if the maximum mass exceeds $2 M_\odot$.
\end{itemize}

\begin{figure}
	\begin{centering}
		\epsfig{file=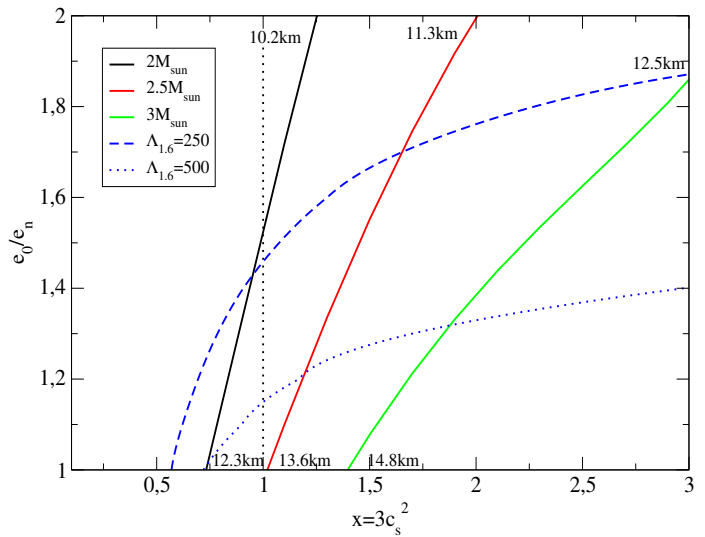,height=9cm,width=7cm,angle=-90}
		\caption{Solid lines: curves of constant maximum mass in the parameter space of the CSS model. At the beginning and at the end of those lines the radii corresponding to the maximum mass configurations are also indicated. The vertical dotted line indicates the conformal limit. The blue dashed and dotted lines are curves of constant $\Lambda_{1.6}$, for the two values $\Lambda_{1.6}= 250$ and $500$, which are close to the $68\%$ and $90\%$ confidence interval obtained from the analysis of GW170817, see Fig.3.}
	\end{centering}
	\label{fig:cont-mmax}
\end{figure}

QSs are self-bound objects that can be thought as giant nuclear drops having a profile which ends at the surface of the star with a finite (and large) value of the energy density $e_0$, of the order of that of nuclear matter at saturation. It is well known that for this kind of objects, 
the maximum mass does depend on the speed of sound, as in the case of ordinary neutron stars, but also on the value of $e_0$.
In particular, when adopting the most simple prescription i.e. the constant speed of sound (CSS) EoS $p=c_s^2(e-e_0)$ where $p$ and $e$ are pressure and energy density \citep{Alford:2013aca,Zdunik:2012dj,Chamel:2012ea,Drago:2019tbs}~\footnote{Notice that this simple prescription provides EoSs very similar to the ones obtained within the MIT bag model in which the role of $e_0$ is played by the bag constant and $c_s^2\sim 1/3$ since the up and down quarks are basically massless while the strange quark is massive but with a value of mass of $\sim 100$ MeV and thus small with respect to the chemical potential, see \cite{Alford:2004pf}. Similarly, also within the NJL model $c_s^2 \sim 1/3$, see \cite{Ranea-Sandoval:2015ldr}.} one can show that the maximum mass 
$m_{\mathrm {\mathrm{\mathrm{max}}}} \propto e_0^{-1/2}$ \citep{Lattimer:2010uk}.
Unfortunately, there exists no such simple scaling with $c_s$. We computed numerically the maximum mass (often called Tolman-Oppenheimer-Volkoff mass) as a function of $c_s$ and $e_0$ and found the following parametrization: 
$m_{\mathrm max}(x,e_0)=(e_n/e_0)^{1/2} (0.11 + 3.27 x - 1.04 x^2 + 0.13 x^3)$ where $x=c_s^2/(1/3)$ 
and $e_n=150$ MeV fm$^{-3}$ is the nuclear matter energy density at saturation.
We display in Fig.~\ref{fig:cont-mmax} the contour plot of $m_{\mathrm{\mathrm{max}}}$. One can notice that in the conformal limit $x=1$, it is possible to reach the range $2M_{\odot}\lesssim  m_{\mathrm{\mathrm{max}}}\lesssim 2.5M_{\odot}$ provided that $1\lesssim e_0/e_n \lesssim 1.5$.
Thus, $c_s$ need not to exceed the conformal limit in order to have masses larger than $2M_{\odot}$ when QSs are considered. 
Actually, one can also notice from the same plot that even a value of $x$ as low as $x \sim 0.75$ allows to reach $m_{\mathrm{\mathrm{max}}} \sim 2M_{\odot}$ provided that $e_0$ coincides with the nuclear matter density $e_n$. While such a low value of $e_0$ is probably unrealistic, this result indicates that there is a window of values for $c_s$ and $e_0$ that allows to obtain massive stars even if $c_s$ is well below the conformal limit.
Indeed, in \cite{Dondi:2016yjl} by adopting the chiral color dielectric model for computing the EoS of quark matter, it has been found that it is possible to obtain values of $m_{\mathrm{\mathrm{max}}}$ close to the $2M_{\odot}$ limit while $c_s$ approaches the conformal limit from below. The existence of two-families of compact stars (CSs) 
removes therefore the tension between astrophysical observations and the 
theoretical bounds on $c_s$ \citep{Bedaque:2014sqa} at least for masses not exceeding $\sim 2.5M_{\odot}$.

A correlated issue concerns the radii of QSs.
One can show that the radius of the maximum mass configuration scales as
$r_{\mathrm{\mathrm{max}}}(x,e_0)=(e_n/e_0)^{1/2} (7.06 + 9.40 x - 3.42 x^2 + 0.47 x^3)$. In Fig.~\ref{fig:cont-mmax}, we indicate the range of values of $r_{\mathrm{max}}$ for fixed values of $m_{\mathrm{max}}$. Clearly, the larger the value of $m_{\mathrm{max}}$, the larger the value of $r_{\mathrm{max}}$ (which reaches $\sim15$km for $e_0/e_n=1$ and $m_{\mathrm{max}}=3M_\odot$). In the two-families scenario it is therefore natural to interpret stars with large radii as massive QSs. This will be the starting point of the discussion on the astrophysical data presented in Sec. ~\ref{data}.

Finally, concerning HSs: as shown in \citep{Bedaque:2014sqa} they can reach  $m_{\mathrm{max}} \sim 1.9M_{\odot}$ without violating the conformal limit.
Actually, the formation of hyperons and delta resonances reduces the value of $c_s$ in hadronic matter with a consequent reduction of the values of the maximum mass and of the radii of HSs. Therefore very compact HSs, having $R_{1.4} \lesssim 11$km \citep{Burgio:2018yix} can exist in the two-families scenario together with very massive  QSs \citep{Drago:2013fsa}, while respecting the conformal limit both in the hadronic and in the quark sector.

The previous simple theoretical analysis establishes which are the possible values of $m_{\mathrm{max}}$ in the two-families scenario (by adopting the CSS model for quark matter). Those predictions
need to be confronted with the available astrophysical data.
It is the aim of this work to obtain the posterior distributions of $e_0$ and $c_s$ by performing a Bayesian analysis on a selected sample of data which are interpreted as QSs within the two-families scenario. In our analysis we have not used the recent data obtained by NICER on MSP J0740+6620 \citep{Riley:2021pdl,Miller:2021qha}, instead the estimate of the radius of an object having a mass $\sim 2.05 M_{\odot}$ (as in the case of MSP J0740+6620 \citep{Cromartie:2019kug,Fonseca:2021wxt}) will be a relevant outcome of our analysis.

Finally, we will discuss 
the possibility that the source of GW190814 \citep{Abbott:2020khf} is a BH-CS system
implying that $m_{\mathrm{max}}$ is larger than $2.5M_{\odot}$.

\section{Selection of the sources}\label{data}

The two-families scenario is based on the coexistence of two classes of CSs. The first one, based on a soft hadronic EoS \citep{Drago:2014oja},
is composed by light and very compact HSs, with $R_{1.4}$ of the order of $(10.5-11)$ km and the second one by massive QSs with larger radii \citep{Drago:2013fsa,Drago:2015cea}.
The formation of new degrees of freedom inside HSs softens the EoS leading to a maximum mass $m_{\mathrm{max}}^{HS}$ of $\sim (1.5 - 1.6) M_{\odot}$.
For a same baryonic mass, QSs have a gravitational mass lower than the one of HSs, of the order of $0.1 M_{\odot}$ or more \citep{DePietri:2019khb} 
and thus the transition from HSs to QSs is strongly exothermic \citep{Berezhiani:2002ks,Bombaci:2004mt,Drago:2004vu,Bombaci:2006cs,Drago:2020gqn}. The process of transformation of a HS into a QS can start only when there are enough hyperons at the center of the HS to form the first stable droplet of strange quark matter which can then trigger the conversion, see \cite{Drago:2005yj,Herzog:2011sn,Pagliara:2013tza,Drago:2015fpa}. 
In turn, this implies the existence of a minimum mass for the QSs branch (if QSs generate from the conversion of a HS), $m_{\mathrm{min}}^{QS}\sim m_{\mathrm{max}}^{HS} - 0.1 M_{\odot}$ and the coexistence of both HSs and QSs in the interval $[m_{\mathrm{min}}^{QS},m_{\mathrm{max}}^{HS}]$. In this range, HSs and QSs can have the same mass but different radii.

We use the simultaneous measurements of masses and radii for several X-ray sources, and also the masses and tidal deformabilities derived from the gravitational wave events reported by LIGO-VIRGO collaboration (LVC) \citep{advanced-ligo,advanced-virgo}. Specifically, for our sample of possible QSs candidates we have chosen 4U 1724-07, SAX J1748.9 2021, 4U 1820--30,  4U 1702--429,  J0437--4715 \citep{Ozel:2015fia,Nattila:2017wtj,Gonzalez-Caniulef:2019wzi}, the high-mass component of GW170817 \citep{TheLIGOScientific:2017qsa,Abbott:2018exr}, and both the components of GW190425 \citep{Abbott:2020uma}. For 4U 1702--429 \citep{Nattila:2017wtj} and J0437--4715 \citep{Gonzalez-Caniulef:2019wzi}, we take a bivariate Gaussian distribution to resemble the $M-R$ posterior since the full distribution is not available. For all the other sources we have instead used directly the posterior distributions which are publicly available.
Our data refer to CSs which can be interpreted, within the two-families scenario, as being QSs, as we explain in the following. 

Let us clarify how we have chosen the sources in our analysis on the EoS of QSs.
First, we classify J0437--4715 as a QS, because its radius is larger than $\sim 13$ km, the central value of its mass is $\sim 1.45 M_{\odot}$ and HSs in the two-families scenario cannot have such a large radius for that value of mass. We will also choose that mass value as a guess for $m_{\mathrm{min}}^{QS}$.
Thus $m_{\mathrm{max}}^{HS}$ is fixed to $m_{\mathrm{min}}^{QS}+0.1M_{\odot}= 1.55M_{\odot}$ and all the sources with masses  $\gtrsim 1.55M_{\odot}$ can be interpreted as QSs. In our analysis the masses have been chosen to be the 
mean values of the marginalized distribution of the sources. This criterion is fulfilled by all the sources which we have selected in this study apart from GW170817\_1, whose mass ($\sim 1.49 M_{\odot}$) falls in the coexistence region $[m_{\mathrm{min}}^{QS},m_{\mathrm{max}}^{HS}]$. 
The reason for assuming that GW170817\_1 is a QS is phenomenological: the presence of strong electromagnetic
counterparts associated with GW170817 (GRB170817A and AT2017gfo) 
implies that the merger event did not produce a prompt collapse.
Since the threshold mass for a prompt collapse for a HS-HS binary has been estimated in \cite{DePietri:2019khb} to be $m_{thr}=2.5 M_{\odot}$, that event cannot be interpreted as a HS-HS merger within the two-families scenario and it must be classified as HS-QS merger.
Thus, the heaviest component,  GW170817\_1, within the two-families scenario, is most probably a QS.

\section{Results of the Bayesian analysis}\label{result}

Let us present the results of our Bayesian analysis on the sources discussed in Section \ref{data}. We have used the CSS parameterized EoS described before, with two free parameters: $e_0$ and $c_s^2$. We select the priors assuming for $e_0$ a flat distribution in the range $(160-232)$ MeV fm$^{-3}$ and for $c_s^2$ a flat distribution in the range [$0.1,1$]. We request $m_{\mathrm{max}}^{QS}> 2.05 M_\odot$, what has obvious consequences on the ranges of the parameters. Indeed, because of that constraint, in \cite{Traversi:2020dho} the prior range for $e_0$ was restricted to be $e_0<220$ MeV fm$^{-3}$ since $c_s^2$ was fixed to $1/3$. 
This restriction does no longer hold in the present two-parameters analysis, since an increase in $c_s^2$ allows for larger values of $e_0$. 
Notice anyway that $c_s^2 >0.26$ in order to reach $2 M_\odot$, as shown in Fig. 1.
Details of this analysis alongside a comparison with the neural network based prediction techniques can be found in \cite{Traversi:2020dho}.

Next, we construct the joint posterior and investigate the conformal limit on $c_s$ given the present data.
We use the publicly available posterior samples of X-ray measurements from \citet{Ozel:2015fia} \footnote{The mass-radius distributions of the sources discussed in \citet{Ozel:2015fia} can be found at \href{http://xtreme.as.arizona.edu/neutronstars/}{http://xtreme.as.arizona.edu/neutronstars/}.}  while the mass and tidal deformability samples of the binary merger components are provided by LVC \footnote{The data from GW170817 and GW190425 are available respectively at \href{https://dcc.ligo.org/LIGO-P1800115/public}{https://dcc.ligo.org/LIGO-P1800115/public} and 
\href{https://dcc.ligo.org/LIGO-P2000026/public}{https://dcc.ligo.org/LIGO-P2000026/public}.}.

\begin{figure}
    \centering
         \includegraphics[width=0.5\textwidth]{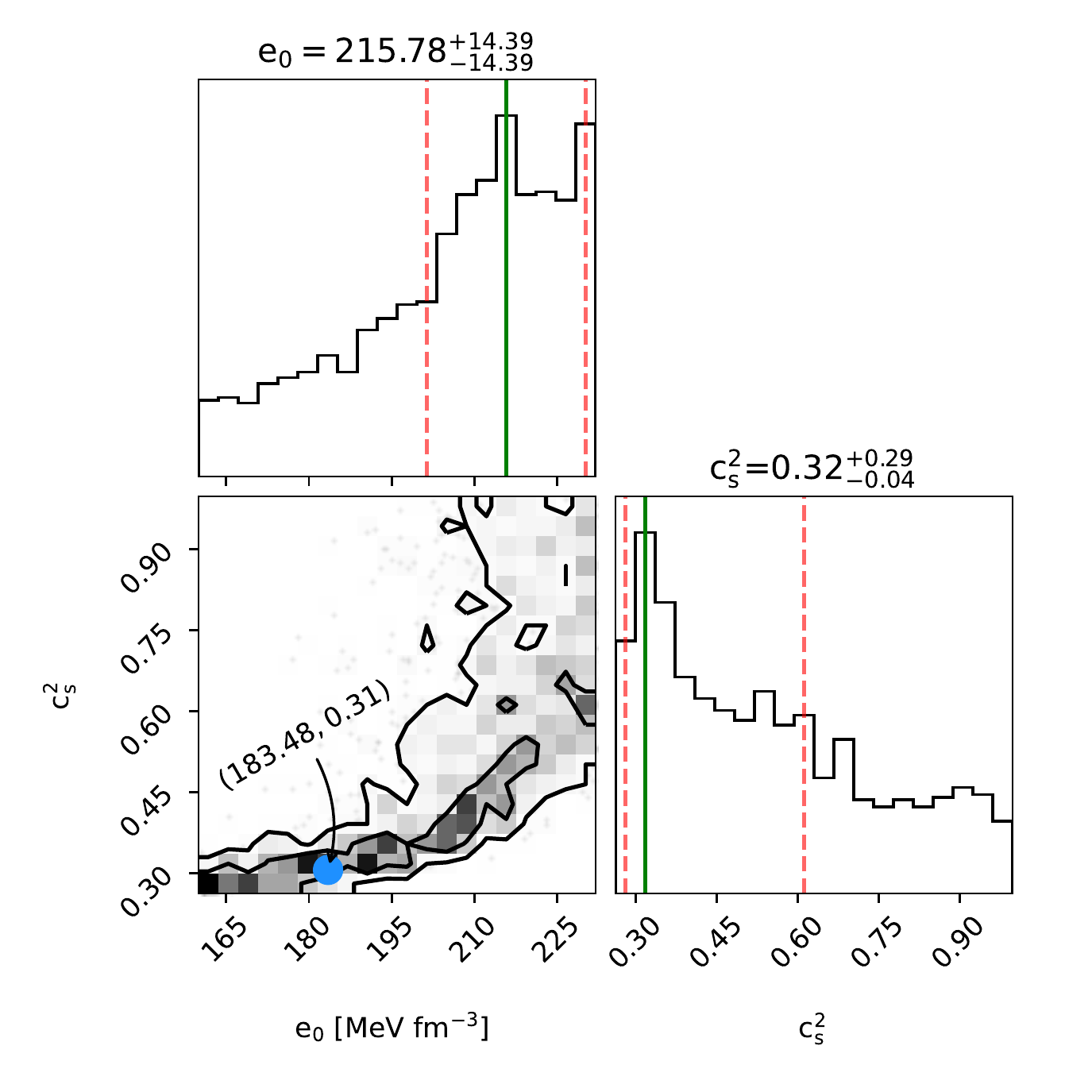}
        \caption{Posterior for $e_0$ and $c_s^2$ resulting from the Bayesian analysis. The blue circle shows the most probable point of the joint distribution. In the correlation plot, the inner and the outer curves correspond to the $1\sigma$ and $90\%$ CI, respectively. The green solid lines correspond to the modes of the marginalized PDFs, while the red dashed lines show the corresponding $1\sigma$ errors. }
    \label{e0_cs}
\end{figure}

\begin{figure*}
    \centering
        \begin{tabular}{cc}
        \includegraphics[width=0.48\textwidth]{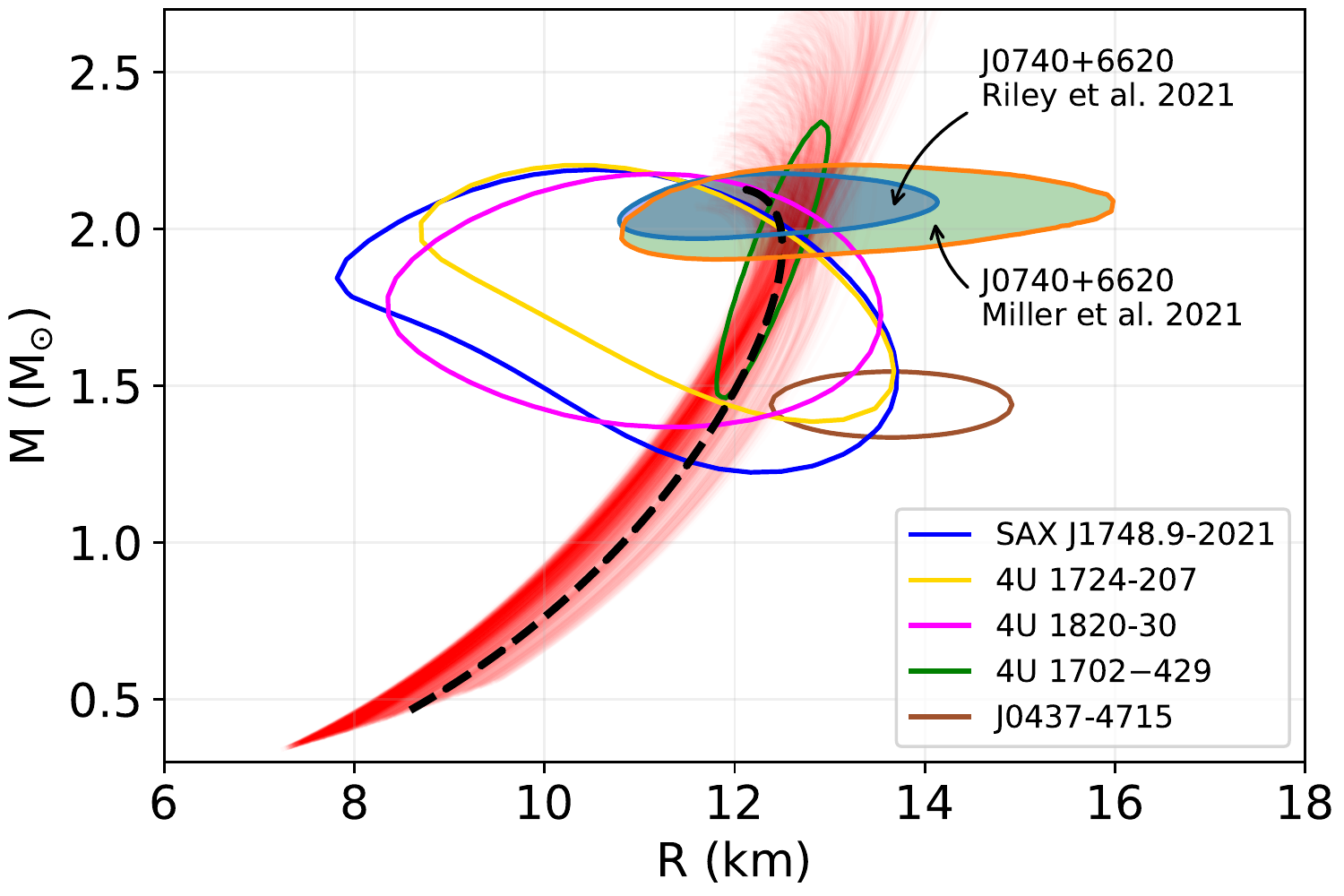}& \includegraphics[width=0.48\textwidth]{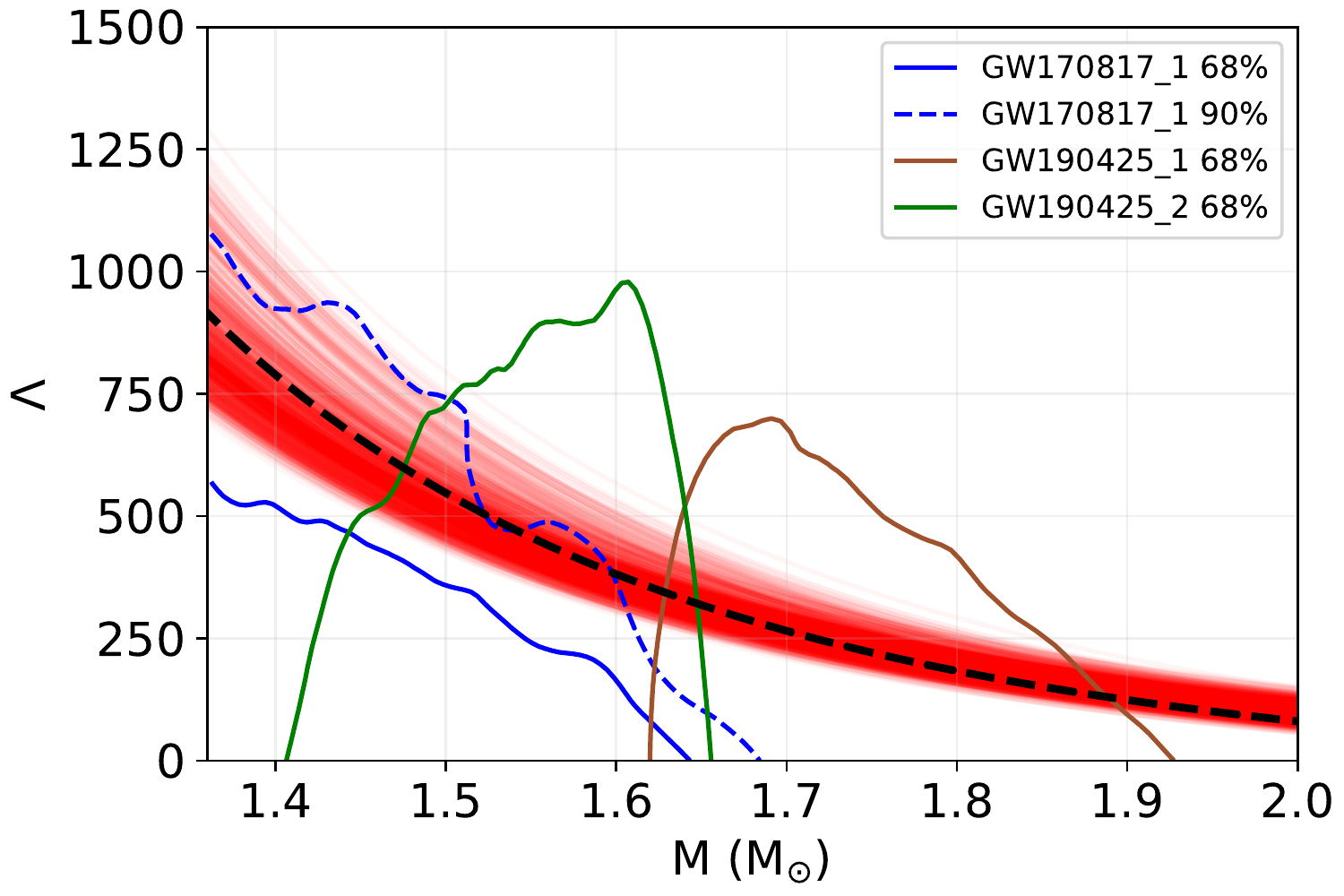}
         \end{tabular}
        \caption{Inferred $M-R$ (left) and $M-\Lambda$ (right) curves for the 68\% CI corresponding to the EOS posterior along with the sources also at 68\%. The light blue and green patches correspond to the latest mass and radius measurements of the pulsar J0740+6620 \citep{Fonseca:2021wxt,zenodo_riley_0740,zenodo_miller_0740}. The black dashed lines represent the most probable EOS. Additionally, the blue dashed curve on the right panel denotes the 90\% of the heavier component of GW170817.}
    \label{curves_bayes_2p}
\end{figure*}

In figure \ref{e0_cs}, we display the marginalized PDFs for $e_0$ and $c_s^2$ along with their most probable values and the $1\sigma$ errors, and the $2D$ distribution with $1\sigma \ (39.3\%)$, and $90\%$ credible interval (CI). The quantiles are not directly useful to our analysis, but they indicate the trends in the posteriors. The region with larger likelihood corresponds to lower values of $c_s^2$ with corresponding not-too-large values of $e_0$. Indeed, a positive correlation among these two parameters is found inside the $1\sigma$ region.
The most probable point of the joint PDF is represented as a blue circle in figure \ref{e0_cs} and it is located at $e_0=183.48$ MeV fm$^{-3}$, $c_s^2 = 0.31$. This is a remarkable result: 
by adopting the two-families scenario and by performing a Bayesian analysis on the QSs candidates, we find that the preferred values of $c_s$ not only are far from the causal limit, but even that violations of the conformal bound seem to be unnecessary. Notice also that this result cannot be reduced to the more trivial information that it is possible to reach large masses with QSs without violating the conformal limit. As shown in Fig. \ref{fig:cont-mmax}, if the observed radii were smaller the violation of the conformal limit would be necessary (at least with the rather trivial CSS EoS adopted here). Instead we have demonstrated that: 1) the conformal limit can be respected with an EoS allowing for the existence of QSs because the radii are rather large; 2) for the observed values of masses and radii the solution respecting the conformal limit is not only possible but it is even the most probable.

The $M-R$ and $M-\Lambda$ curves corresponding to the 68\% CI are shown in figure \ref{curves_bayes_2p}. The higher likelihood region of the $2D$ posterior is associated to EoSs which are characterized by $m_{\mathrm{max}}$ in the range $ \sim (2.1-2.2) M_{\odot}$ and radii in the interval $R_{1.6}\sim (11.6-12.9)$ km.
We obtained for the most probable solution $m_{\mathrm{max}}=2.13 M_{\odot}$ and $R_{1.6}= 12.20$ km. The corresponding curves in both panels in Fig. \ref{curves_bayes_2p} are represented with a black dashed line. While the present work was in progress, the mass (and the distance) of J0740+6620 was measured again by using radio data \citep{Fonseca:2021wxt} and X-ray observations by NICER have constrained its mass-radius relation \citep{Riley:2021pdl,Miller:2021qha}. These latest results are not included in our analysis, but they are shown in figure \ref{curves_bayes_2p} together with the outcome of our Bayesian analysis and it is possible to appreciate that our results are completely consistent with these new findings. Concerning the three tidal deformabilities of the stars included in our sample, it is clear that while there is a good agreement with the two components of GW190425, for GW170817\_1 there is no overlap between the theoretical curves and the posterior distribution at the
$68\%$ CI. However at $90\%$ CI the agreement is met also with GW170817\_1 (see the blue dashed line of Fig.~\ref{curves_bayes_2p}, right panel). We believe that a more sophisticated EoS can probably solve this (mild) tension.

\section{What if GW190814 has a $2.6M_{\odot}$ compact star component? }
The detection of the gravitational waves signal GW190814 by the LVC collaboration \citep{Abbott:2020khf} led to the discovery of a binary system made of a $23M_{\odot}$ BH and a CS companion of $\sim 2.6M_{\odot}$ (the lower value at $90\%$ credible level being $2.5M_{\odot}$). The nature of the companion is rather uncertain since its mass falls within the BH lower mass gap but, on the other hand, such a massive neutron star
challenges previous astronomical constraints and nuclear physics constraints \citep{Fattoyev:2020cws}. In the recent work by \cite{Bombaci:2020vgw}, it has been proposed that the companion could actually be a QS.

\begin{figure}
	\begin{centering}
		\epsfig{file=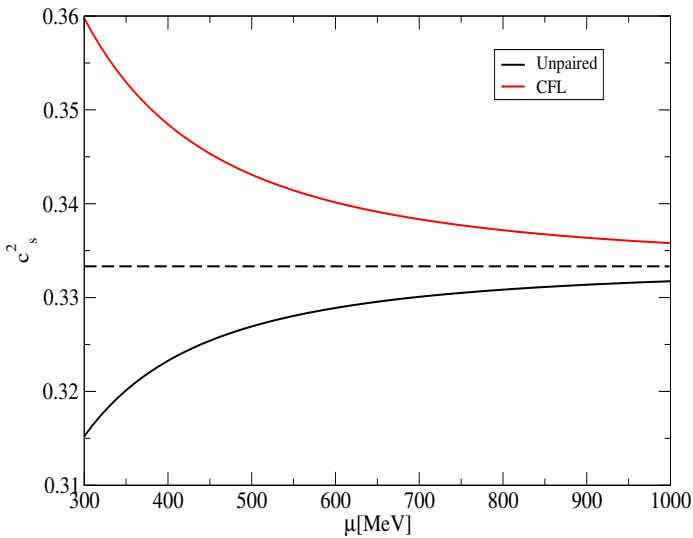,height=9cm,width=7cm,angle=-90}
		\caption{Speed of sound as a function of the quark chemical potential $\mu$ for unpaired quark matter and CFL matter. For the latter we use $m_s=100$, $\Delta=80$ MeV, $B^{1/4}=135$ MeV and the pQCD correction parameter is $a_4=0.7$, see text.}
		\label{csmu}
	\end{centering}
\end{figure}

From the discussion presented in Sec.~\ref{intro} it is clear that to explain the existence of a QS with $M=2.6M_{\odot}$ the conformal limit must be violated also in dense quark matter. Namely, it is necessary that for a certain range of densities $c_s^2$ is larger than $1/3$. Alternatively,
one should find a physical mechanism that reduces the value of $e_0$.
In \cite{Bombaci:2020vgw}
this requirement is achieved by assuming that quark matter could be in a color superconducting state, namely the CFL phase. The pressure $P$ in this case reads: $P=3/(4 \pi^2)a_4\mu^4-3/(4 \pi^2)(m_s^2-4\Delta^2)\mu^2-B$,
where $\Delta$ is the superconducting gap, $\mu$ the quark chemical potential and $m_s$ the strange quark mass \cite{Alford:2004pf}. The additional term, proportional to $m_s^2-4\Delta^2$, introduces a density dependence of $c_s$, as shown in figure~\ref{csmu}, and it also modifies the value of $e_0$. 
In particular, if $\Delta>m_s/2$, $c_s$ approaches the conformal limit from above and the value of $e_0$ is reduced.
In this way it is possible to reach values of $m_{\mathrm{max}}$ larger than $2.5 M_{\odot}$ as shown in \cite{Bombaci:2020vgw} and to pin down a physical mechanism for explaining the violation of the conformal bound at low densities. Notice anyway that the deviation from the conformal limit is rather small. Actually, the most important effect of the superconducting gap on the value of the maximum mass 
derives from the reduction of $e_0$ which, in this numerical example, is only slightly larger than $e_n$, namely $e_0 =152$ MeV fm$^{-3}$. For the same set of parameters (see figure caption), but with $\Delta=0$, one obtains $e_0=190$ MeV fm$^{-3}$. 

In conclusion, one can ask to what extent deviations from the conformal limit, within the two-families scenario, are necessary in the case in which $m_{\mathrm{max}}>2.5M_{\odot}$. It clearly depends on how compact or (tidal deformable) QSs are. In Fig. 1, we display the $68\%$ and $90\%$ credible interval for the tidal deformability $\Lambda$ of a $1.6M_{\odot}$ star obtained from the analysis of GW170817: $\Lambda_{1.6} \lesssim 250$ and $\Lambda_{1.6} \lesssim 500$, see also Fig. 3 right panel. Those constraints can be satisfied if $1.2 \lesssim x \lesssim 1.6$, a violation of the conformal limit which is anyway significantly smaller than the one needed in the case of hybrid stars, see \cite{Xie:2020rwg}.

\section{Conclusions}
We have argued that the existence of massive compact stars, with masses up to $2.5 M_{\odot}$ does not necessarily imply the violation of the conformal limit in dense strongly interacting matter, in agreement with the results of \cite{Lattimer:2010uk}. One has however to abandon the assumption that only one family of compact stars exists and assume that two separated branches in the mass-radius plot are possible: the HS branch and the QS branch. Within the two-families scenario, some astrophysical sources can be (with a certain degree of uncertainty) identified as QSs. A Bayesian analysis on such selected sample of sources has allowed to estimate the posterior distributions of the two parameters of the model for the EoS, namely $e_0$ and $c_s$. Interestingly, the presently available data suggest that the distribution of $c_s^2$ is actually peaked at a value close to $1/3$.
Our result is therefore totally different from the case of hybrid stars (namely stars with a quark matter core), for which a recent Bayesian analysis has shown that $c_s^2$ has a distribution peaked at $0.95$ \citep{Xie:2020rwg}.

It is quite remarkable that
the most probable equation of state obtained through our Bayesian analysis predicts a radius for PSR J0740+6620 which falls within the recent limits found by NICER \citep{Riley:2021pdl,Miller:2021qha}. Also interesting is that the maximum mass for a QS in our most probable case (which does not violate the conformal limit) is very close to the limits obtained by studying GW170817 and the associated kilonova at2017gfo \citep{Margalit:2017dij,Rezzolla:2017aly}. Moreover, a recent analysis of the maximum mass \cite{Shao:2020bzt} suggests that $m_{\mathrm{max}}= 2.26 (+0.12/-0.05) M_\odot$. In the normal scenario based on HSs, masses even slightly larger than 2 $M_\odot$ would imply a violation of the conformal limit and therefore, if the conformal limit has to be respected at all densities, then (at least) the most massive stars need to be QSs.

One can compare our result with e.g. the one obtained in \cite{Thapa:2021kfo}. In both cases the new limits obtained by NICER are respected and in both cases the speed of sound does not violate (at least significantly) the conformal limit. In the scenario described in \cite{Thapa:2021kfo} the speed of sound is reduced by the production of resonances, hyperons and kaon condensation. However in that way masses significantly larger than 2 $M_\odot$ cannot be obtained and the radius of a $1.4 M_\odot$ star must be $R_{1.4}\gtrsim 11.5$km. Instead, in the two-families scenario the most massive stars are QSs and large masses up to $\sim 2.5 M_\odot$ can be obtained without the need of violating the conformal limit. Notice that in the two-families scenario the first family is also composed of nucleons, resonances and hyperons, but that family need not to reach large masses and $R_{1.4}$ can be even significantly smaller than 11 km. Both scenarios are "realistic", since they do not artificially suppress the production of resonances and hyperons and future astrophysical data will most likely be able to test their predictions.

If compact stars with masses larger than $2.5M_{\odot}$ will be discovered, then even in the two-families scenario the violation of the conformal limit is mandatory. A possible mechanism has been suggested which is based on the formation of a (sizable) superconducting gap. By using the constraints on the tidal deformability derived from GW170817 we found that these deviations need not to exceed $\sim 60\%$ and are thus significantly smaller than in the case of hybrid stars investigated in \cite{Xie:2020rwg}.

Finally, while in this paper the Bayesian analysis has been performed by using 
the simple constant speed of sound model, it would be important in future investigations to use more sophisticated models allowing for a density dependence of $c_s$ such as e.g. the color dielectric model of \cite{Dondi:2016yjl} or a simple parameterization with piecewise constant speed of sound models.

\begin{acknowledgements}
 PC acknowledges past support from INFN postdoctoral fellowship. PC is currently supported by the Fonds de la Recherche Scientifique-FNRS, Belgium, under grant No. 4.4503.19.
\end{acknowledgements}
 
\bibliographystyle{aa}
\bibliography{mybiblio}
\end{document}